# 4K-Memristor Analog-Grade Passive Crossbar Circuit

H. Kim[*,†], H. Nili[†], M. R. Mahmoodi, and D. B. Strukov[*#]

Department of Electrical and Computer Engineering, University of California, Santa Barbara,
CA 93106, USA

The superior density of passive analog-grade memristive crossbars may enable storing large synaptic weight matrices directly on specialized neuromorphic chips, thus avoiding costly off-chip communication. To ensure efficient use of such crossbars in neuromorphic computing circuits, variations of current-voltage characteristics of crosspoint devices must be substantially lower than those of memory cells with select transistors. Apparently, this requirement explains why there were so few demonstrations of neuromorphic system prototypes using passive crossbars. Here we report a 64×64 passive metal-oxide memristor crossbar circuit with ~99% device yield, based on a foundry-compatible fabrication process featuring etch-down patterning and low-temperature budget, conducive to vertical integration. The achieved ~26% variations of switching voltages of our devices were sufficient for programming 4K-pixel gray-scale patterns with an average tuning error smaller than 4%. The analog properties were further verified by experimentally demonstrating MNIST pattern classification with a fidelity close to the software-modeled limit for a network of this size, with an ~1% average error of import of ex-situ-calculated synaptic weights. We believe that our work is a significant improvement over the state-of-the-art passive crossbar memories in both complexity and analog properties.

## Introduction

Analog-grade nonvoltaile memories, such as those based on floating-gate transistor [1-3], and phase-change [4-6], ferroelectric [7, 8], magnetic [9], solid-state electrolyte [10, 11], organic [12], and metal-oxide [13-21] materials, are enabling components for mixed-signal circuits implementing vector-by-martix multiplicaiton (VMM), which is the most common operation in any artificial neural network. The main advantage of using passively integrated metal-oxide memristors [22], which are also commonly referred as RRAMs or ReRAMs [23-25], in such circuits are their superior density and lower fabrication cost [25]. In fact, due to excellent scaling prospects and analog properties, vertically-integrated ReRAMs might challenge in effective density much slower 3D NAND memories [26]. The main downside of such circuits is lower input / output impedance, which leads to larger VMM peripheral overhead, especially for memory devices with larger conductances. However, such overhead can be reduced via efficient sharing of peripheral resources, at the cost of decreasing computational throughput [27, 28], which should be acceptable given the very high speed of memristor-based VMMs [29].

---
[†] These authors contributed equally to this work. H. Kim is now with the Department of Electronic Engineering in Yeungnam University, Gyeongsan 38541, Korea. Corresponding authors: *hyungjin@yu.ac.kr, #strukov@ece.ucsb.edu .

High integration density would be essential for enabling hardware implementations of large neural network models, e.g. used for end-to-end automatic speech recognition [30] and natural language translation [31], on a single chip without having to perform very energy-taxing and slow data transfer with the off-chip memory. For example, the largest multilingual neural model for automatic translation among seven common languages feature 640 million parameters [31]. Furthermore, a more advanced mixture-of-expert networks with up to 137 billion parameters have been recently suggested to improve functional performance of language modeling [32]. Assuming 25% memory array efficiency in the most practical, general purpose neuromorphic computing chip [28] (with the remaining 75% of area devoted to array periphery and other functions) storing that many parameters on a chip would require terabit-scale memories, which could be hardly met with planar embedded memory technologies. Though, the complexity of the mentioned networks is expected to reduce with further improvements in algorithms, it is clear that extremely large models would be still very useful. This can be indirectly evidenced by the human brain. With its $\sim 10^{15}$ synapses, it can serve as a proxy for the complexity of the future highly cognitive neuromorphic systems [33].

There has been substantial progress in the development of 1T1R ReRAM arrays in which memory cell is coupled with dedicated select transistor [25, 34], and many demonstrations from academia and industry of using such active memories in neuromorphic computing circuits – see, e.g., [18-20] and also recent reviews [35-38]. However, the progress in the most prospective, passive analog-grade ReRAM circuits has been much slower [14, 15, 17, 21], mainly because of much harsher requirements for the uniformity of memory cells' *I-V* characteristics [39]. For example, Xpoint memory - the most advanced commercialized technology using passively-integrated memory devices - operates in a digital mode. (Such memory is also most likely based on phase-change materials [40] which are less appealing for analog computing applications due to larger conductance drift over time.) A promising *I-V* uniformity results with very tight variations were reported for stand-alone devices based on epitaxial [41] and organic [42] materials, though the main concern is compatibility of the utilized fabrication flows with conventional semiconductor processes. Ref. 21 reported a detailed switching and conductance tuning statistics for 250-nm half-pitch 20×20 crossbar circuits with $Pt/Al_2O_3/TiO_{2-x}/Ti/Pt$ devices and presented results for mixed-signal multi-layer perceptron based on such circuits. Sparse encoding algorithm was experimentally realized in 500-nm half-pitch 32×32 circuits based on $W/WO_x/Pd/Au$ devices [17]. In these works, the reported standard variation for the tuning precision were 7.5% and >25%, respectively. (The latter number is estimated based on Fig. S3d data [17]. It is not clear, however, if such data were calculated based on reading conductances after completing tuning process for all devices in the crossbar or, alternatively, just a fraction of them, as it was performed by the same authors in Ref. 43. This is important issue considering that the main factor degrading tuning accuracy in passively integrated circuits is half-select disturbance [28, 39], which, e.g., is not a problem for 1T1R memory arrays.)

The complexity, uniformity, and analog properties of previously reported works on memristive crossbar circuits are clearly not sufficient for practical use of this technology in neuromorphic applications. The main contribution of this work is to address these challenges by developing uniform fabrication technology for building larger 3D crossbar array circuits and to show potentials of such technology in neuromorphic computing applications. The developed circuits have more than ten times more devices as compared to previous work [21] which reported similarly detailed characterization statistics. Moreover, the demonstrated artificial neural network is close in complexity to the state-of-the-art neuromorphic prototypes based on (much sparser) 1T1R ReRAM devices [38].

**Device Fabrication**

The developed 64×64 crossbar circuit consists of Ti/Al/TiN top and bottom electrodes and $Al_2O_3/TiO_{2-x}$ switching layer (Figs. 1a and S1). The actual crossbar array dimensions are (64+2)×(64+2), with an additional line added at both sides of the circuit for the top and bottom layers to achieve better uniformity for the devices in the main array. Specifically, the first step in the fabrication was deposition of Ti (10 nm) / Al (70 nm) / TiN (45 nm) metal stack on 4-inch Si wafer with 200 nm of thermally grown $SiO_2$ by means of reactive sputtering (Fig. S1a, b). ~250-nm wide bottom electrodes were then patterned by deep ultraviolet lithography stepper with an antireflective coating (Brewer Science DUV-42P), using negative photoresist (Dow Chemical UV2300-0.5) and inductively couple plasma etching process with $BCl_2/Cl_2/N_2$ chemistry to suppress sidewall re-deposition during etching (Fig. S1c). The bottom electrodes were planarized by first depositing 300 nm of $SiO_2$ via plasma-enhanced chemical vapor deposition (Fig. S1d) and then using combination of chemical-mechanical polishing (CMP) process and etching-down with $CHF_3$ plasma (Fig. S1d). The $Al_2O_3$ (1.5 nm) and $TiO_{2-x}$ (30 nm) of the active switching bilayer were deposited, respectively, through atomic layer deposition and reactive sputtering (Fig. S1f). No oxygen descum was conducted after switching layer deposition to keep $TiO_{2-x}$ stoichiometry. ~ 250-nm top electrode lines with Ti (15 nm) / Al (90 nm) / TiN (80 nm) were deposited and patterned similarly to the bottom electrodes (Fig. S1g, h). The switching layer outside the crossbar region was etched with CHF3 plasma to suppress line-to-line leakages and to open ends of bottom electrodes. Ti (40 nm) / Au (400 nm) pads were formed for wire bonding and packaging. Finally, rapid thermal annealing at 350° C in $N_2$ gas with 2% $H_2$ for 1 minute was performed after device fabrication is completed.

Though the developed technology builds upon the previous work on such devices, there are several important improvements that enabled demonstration of functional larger-scale crossbar circuits. The similarities are, for example, in that thin titanium layer in the electrodes provided adhesion, and, in case of the top electrodes, was used to create ohmic interface with large oxygen vacancy concentration near the top portion of titanium oxide film [44]. Instead of relying on precisely controlling stoichiometry during deposition [45], we have opted to thermal annealing to adjust oxygen vacancy profile, which result in more gettering of top electrode titanium metal and diffusion of oxygen vacancies towards the bottom interface [14]. Such fine-tuning of oxygen vacancy doping allowed lowering as-fabricated memristors' conductances and hence reducing voltages for the device forming (and completely eliminating forming process for some), and was necessary for crossbar integration and improving device uniformity. The aluminum oxide layer, with parameters optimized similarly to Refs. [14], is integrated in the stack to suppress device currents at small voltages (and hence improve dynamic range of adjustable currents) and bottom-to-bottom line leakages.

Let us stress the importance of several distinctive already-mentioned techniques which were used to improve line conductances, device uniformity, and yield, and were essential for scaling up the crossbar size. First, aluminum was selected for its better conductivity, instead of commonly used noble and other higher-resistance inert materials in other works [14, 15, 17, 21]. The inert titanium nitride capping was needed to avoid aluminum oxidation. Second, patterning via reactive ion etching instead of typically employed lift-off process allowed fabricating larger (>1/2) aspect-ratio electrodes. Third, the use of CMP greatly improved the quality of top electrodes and, in part, was needed to avoid sidewall residue along bottom line edges (Fig. S2). (This problem is very similar to gate spacer residue at Si fin channel during FinFET process flow [46].) Patterning by etching and CMP also eliminated undesirable formation of kinks at line edges ("rabbit ears") which are typical for lift-off process. (Note that though ion milling and CMP techniques have been previously used to fabricate 2×10×10 crossbar circuits [47], the main purpose of these techniques in that work was to enable vertical monolithic integration, and, e.g., line resistance remained large due to the use of small-aspect ratio Pt electrodes.) With all the modifications, the developed fabrication process has low temperature fabrication budget and can be adopted by silicon foundries for back-end-of-line integration.

Scanning electron microscopy images of the fabricated crossbar array (Figs. 1a, S1i-k) confirms smooth planar topology of the top electrodes and their structural isolation with no noticeable sidewall residue between them.

**Device Characterization**

The devices were formed with positive voltage applied to the top electrode while floating all unselected lines in the crossbar [14]. Because of more extensive annealing compared to previous work [21], the forming voltages and as-fabricated (virgin state) currents (Fig. 1b) were just slightly less compared to, respectively, switching voltages and the device smallest (OFF) state current after forming (Fig. 1c) – see, e.g., the highlighted curves for some specific device. The formed devices show similar in magnitude set and reset voltages (Fig. 2), from 200 µA to 400 µA reset and set switching currents, 2-µA-to-50-µA dynamic current range at 0.25 V with balanced $I(V) \approx I(-V)$ at small voltages, and ~2 at 1 V and ~1.1 (for on state) / ~1.3 (for off state) at 0.25 V average nonlinearities, defined as $0.5 \times I(V)/I(V/2)$.

Fig. 2a shows measured switching dynamics characteristics for all the devices in the 64×64 array. These data were obtained by first setting the conductance of each device to 14 µS with 10% precision. Next, 1-ms-long increasing amplitude pulses with 50 mV step were applied to the device. The device's conductance was read between each programming pulse at 0.25 V, and the sequence of pulses was stopped once the small-voltage conductance exceeded 50 µS. After that, a similar reset/read pulse sequence was applied until the conductance is switched back to 14 µS. The raw experimental data were used to extracted switching thresholds (Fig. 2b-d) which are defined as the smallest voltages (i.e. the specific amplitude of a pulse) at which the conductance of the device changes by more than 20% compared to its initial state.

According to Fig. 2b, the average set and reset threshold voltages are 1.19 V and –1.39 V, respectively, with the standard deviations of 0.31 V and 0.37 V. Though the corresponding normalized variations of ~26% is larger than ~15% and ~20% average spreads in normalized set and reset threshold voltages reported in Refs. 21 and 47, the latter numbers were reported for crossbar circuits with 10× and 20× devices, respectively. Furthermore, there are only 45 (~1.125%) unswitchable devices in the whole crossbar array. The threshold maps show that faulty devices are distributed throughout the array and not contributed by faulty lines but rather stand-alone defects. Also, there is a significant correlation between set and reset voltages, with the devices having larger set voltage likely featuring also larger reset voltage (Fig. S3).

The analog properties of the memristive crossbar were tested by tuning device conductances using fine-tuning algorithm [48]. Such algorithm, which is similar to incremental step programming of flash memory devices [49], is based on applying a sequence of read and write pulses, with a sign and amplitude of write pulses, with up to 2.5 V maximum amplitude and 4 mV / 8 mV of incremental step for set / reset, are adjusted dynamically based on the measured conductance at read pulses. To reduce disturbance of already tuned half-selected devices in passively integrated crossbar circuits, half-biasing scheme was adopted when applying write pulses [14, 15]. Furthermore, to correct a minor conductance drift in some half-selected devices upon programming, tuning of the whole crossbar was performed in several cycles, such that, e.g., all of the devices are tuned, one by one, in the first cycle, and then those which got disturbed are re-tuned in the following cycle. Fig. 3a shows the map of target conductances, representing the grayscale image of Albert Einstein mapped on all devices in 64×64 crossbar array, while Fig. 3b shows their final values after 3 cycles of tuning. The corresponding statistics for the absolute tuning error is plotted in Fig. 3d. Excluding unswitchable devices, for which the error is more than 95%, ~98% of the devices are tuned within 5% error, while the average error is ~3.76%, which is 2× better compared to Ref. 21.

Note that in this experiment, the tuning algorithm was stopped once the desired 5% tuning error was reached, though setting conductances with even higher precision should be possible. This possibility is indirectly indicated by the shape of the tuning error histogram and was experimentally verified by successful conductance tuning of a specific crossbar device with 1% error to linearly spaced values within the lower half of the dynamic range (Fig. 3d), and also by tuning a portion of the crossbar in the context of the demonstrated neuromorphic application – see Fig. 4a-c and its discussion below.

**MNIST Benchmark Classification**

To further verify the analog properties of the crossbar-integrated devices, we have implemented ex-situ trained image classifier and tested it on the common MNIST handwritten digit benchmark [50] (Fig. 4). The considered network is a single-layer perceptron with 64 inputs and 10 outputs. In our experiment, we focused on demonstrating vector-by-matrix multiplication, the core operation in any neural network, while the functionality of neurons, including its bias, was emulated in the software. In particular, the original binary 28×28 MNIST images were down-sampled to 8×8 patterns, so that they can be represented by 64-bit digital vectors in which black/white pixels are encoded by 0 V / 0.25 V voltages and applied to the crossbar vertical lines. By encoding network weights with the corresponding memristor conductances in the 64×10 portion of the crossbar, the currents measured at the virtually grounded horizontal lines of the crossbar represent the results of vector-by-matrix multiplication operation, while the output with the largest current identifies the determined class of the input pattern (Fig. 4a).

The classifier weights were calculated by training the network in software (see Fig. S4 and its caption for more details) and then mapped to the corresponding conductances in the 10-μS-to-100-μS range (Fig. 4b, c). The measured classifier fidelity was compared to the software-based performance of the same network across 1%-to-50% range of weight import errors (Fig. 4d). The results show that experimental data matches closely simulation results and by only 1.87% behind ideal software model for the most accurate weight import.

Fig. 4e, f provides more details on the measured data for the two representative MNIST patterns. Specifically, the first examples show the results of correct classification of pattern "7", with the largest current measured at the 7-th row of the crossbar (Fig. 4e). One the other hand, pattern "9" in the second example is misclassified (Fig. 4f). This is in part because of a large tuning error at unswitchable memristors - see stuck at high-resistance state devices at (9, 22) and (9, 24) locations in the crossbar in Fig. 4c (and also Fig. 2c, d), but also due to narrow current margins between the correct class and the two closest classes representing digits "0" and "8", which is natural given that correct classification in this case would be hard even for a human.

**Summary**

In summary, the general focus of this work was on reducing spread in current-voltage characteristics of integrated metal-oxide memristors, one of the most important problems prohibiting practical use of this technology in neuromorphic computing applications. Our specific contributions include the development of uniform 64×64 passive crossbar circuits with almost 99% working crosspoint metal-oxide memristors based on foundry-compatible fabrication process suitable for back-end-of-line / 3D integration, and experimental demonstrations of conductance tuning with < 4% average error when programming 4K gray-scale pattern and of a single perceptron network implementing ex-situ trained classification of down-sampled 8×8 MNIST benchmark with classification fidelity within 2% of the software-based model. We believe that by showing a 2× smaller tuning error in a crossbar with 10× more integrated devices than in any published work we are aware of, our work is a significant step in the development of passive crossbar memories. The future work should focus on the back-end-of-line integration with conventional semiconductor circuits and decreasing operating and write current of memristors to achieve the ultimate performance promised by this device technology. Previous work showed that the cell currents can be decreased by reducing leakages within the device and between neighboring lines, which can be achieved by patterning active switching layer and scaling down feature sizes [21].

**Methods**

After forming Au pad at each electrode line, the crossbar array was wire-bonded and inserted in a custom printer board for testing and application demonstration (inset in Fig. 1a). The custom-printed circuit board was connected to Keysight tools and controlled by computer setup. Specifically, all electrical measurements were performed using Keysight B1500A parameter analyzer. The connections to crossbar inputs/outputs were steered by Keysight 34980A switching matrix. The parameter analyzer and the switching matrix were connected to a personal computer via general purpose interface and universal serial buses and controlled using a custom C code.

**Acknowledgments**

This work was supported in part by a Semiconductor Research Corporation (SRC) funded JUMP CRISP center, and in part by NSF/SRC E2CDA grant 1740352. Authors thank M. Prezioso for useful discussions.

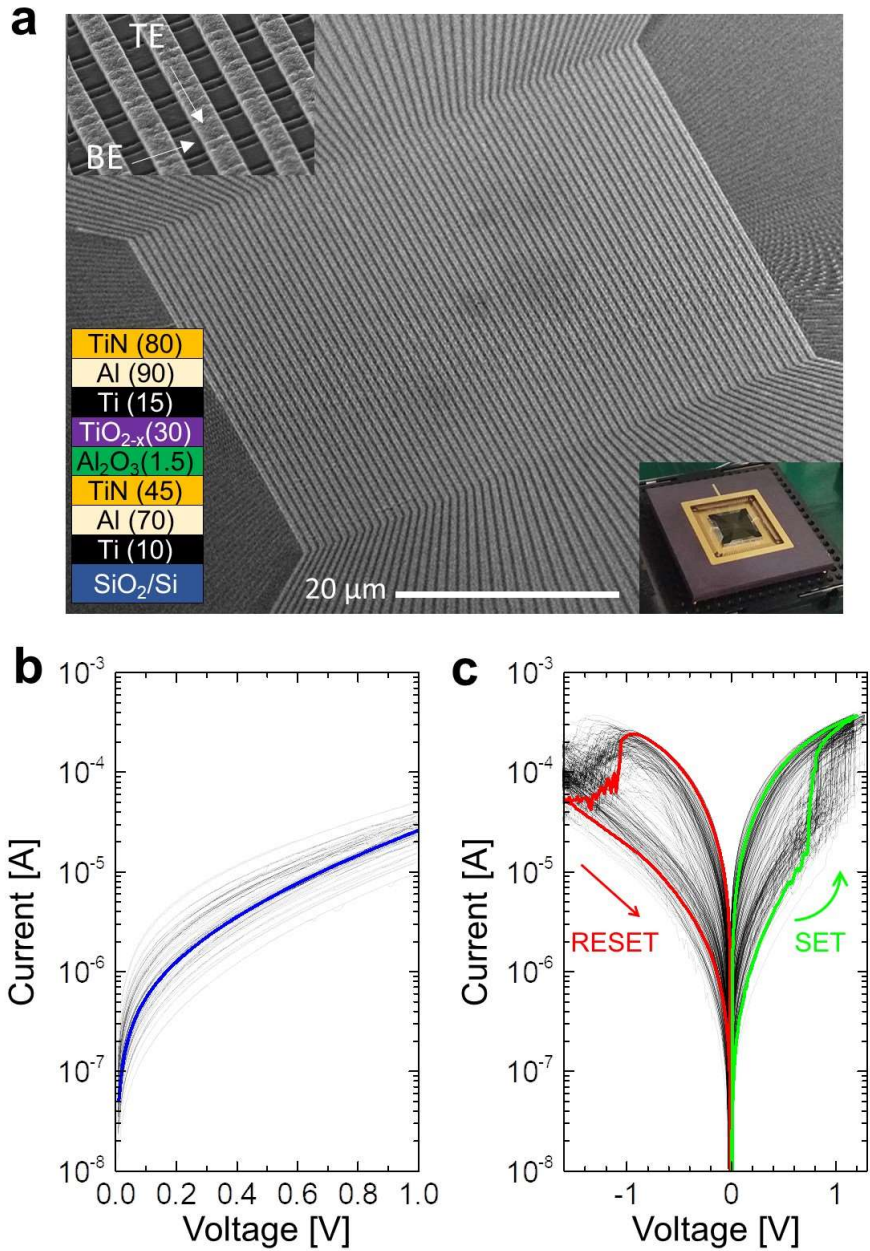

**Figure 1. Memristive crossbar array.** (a) Scanning electron microscope image of the fabricated 64×64 memristor crossbar array. Top, bottom left, and bottom right insets show, correspondingly, zoom-in on the portion of the crossbar, layers at the device cross-section with corresponding thicknesses in nanometers, and packaged chip. (b) Representative as-fabricated and (c) after forming *I-V*s, measured with a quasi-static DC voltage sweeps, shown for the 36 devices of the 6×6 subarray located in the center of the crossbar. For clarity, the curves for a one particular device are highlighted.

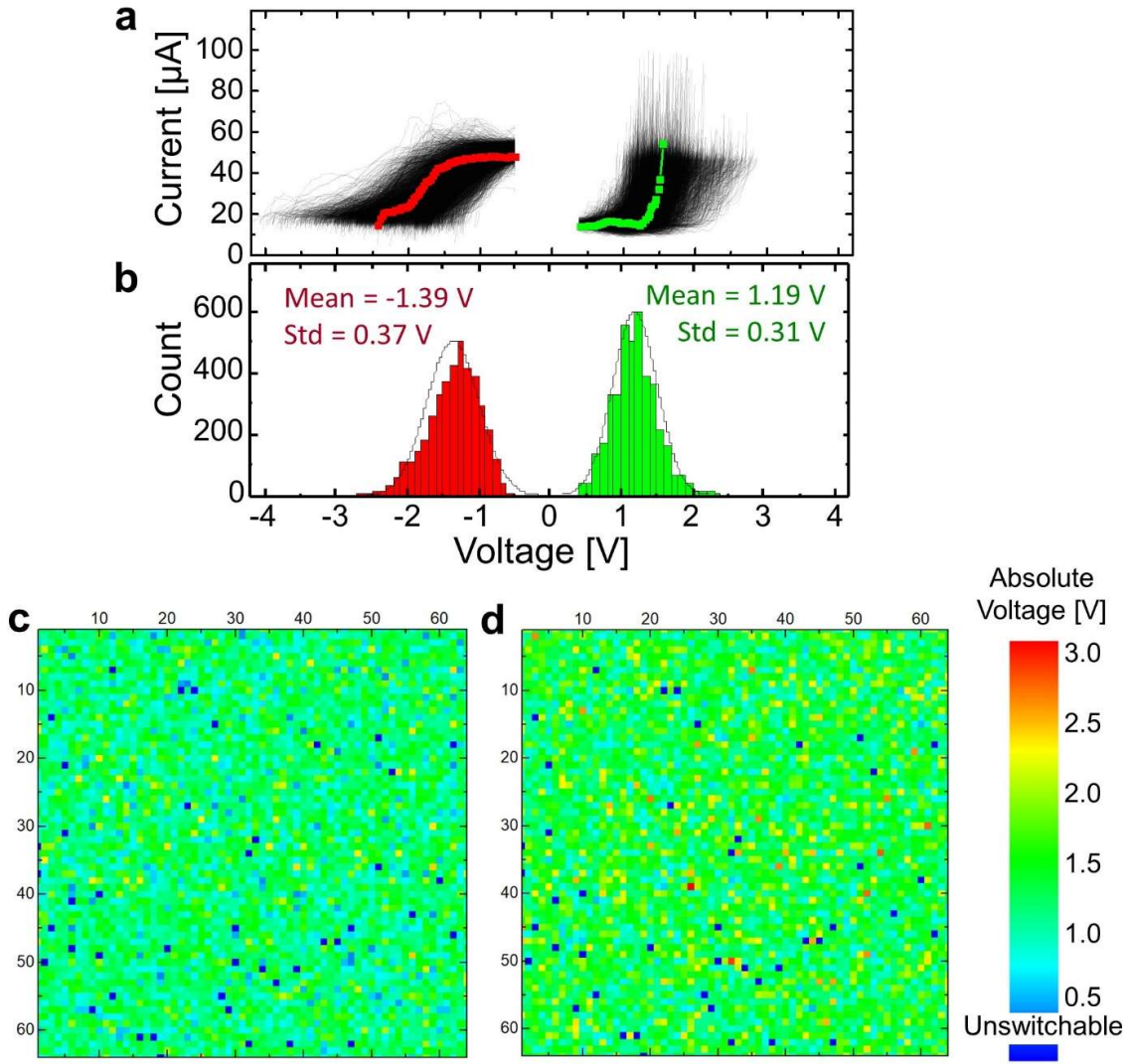

**Figure 2. Switching statistics for crossbar devices.** (a) Measured evolution of conductance upon application of increasing amplitude voltage pulses. All parameters of the utilized pulse sequences are similar to those shown in inset of Fig. 3c, except for 50 mV incremental step. (b-d) Extracted statistics of switching threshold voltages, defined as a smallest absolute voltage at which device conductance measured at 0.25 V changes by 20 %, shown as (b) histogram and voltage map for (c) set and (d) reset transitions.

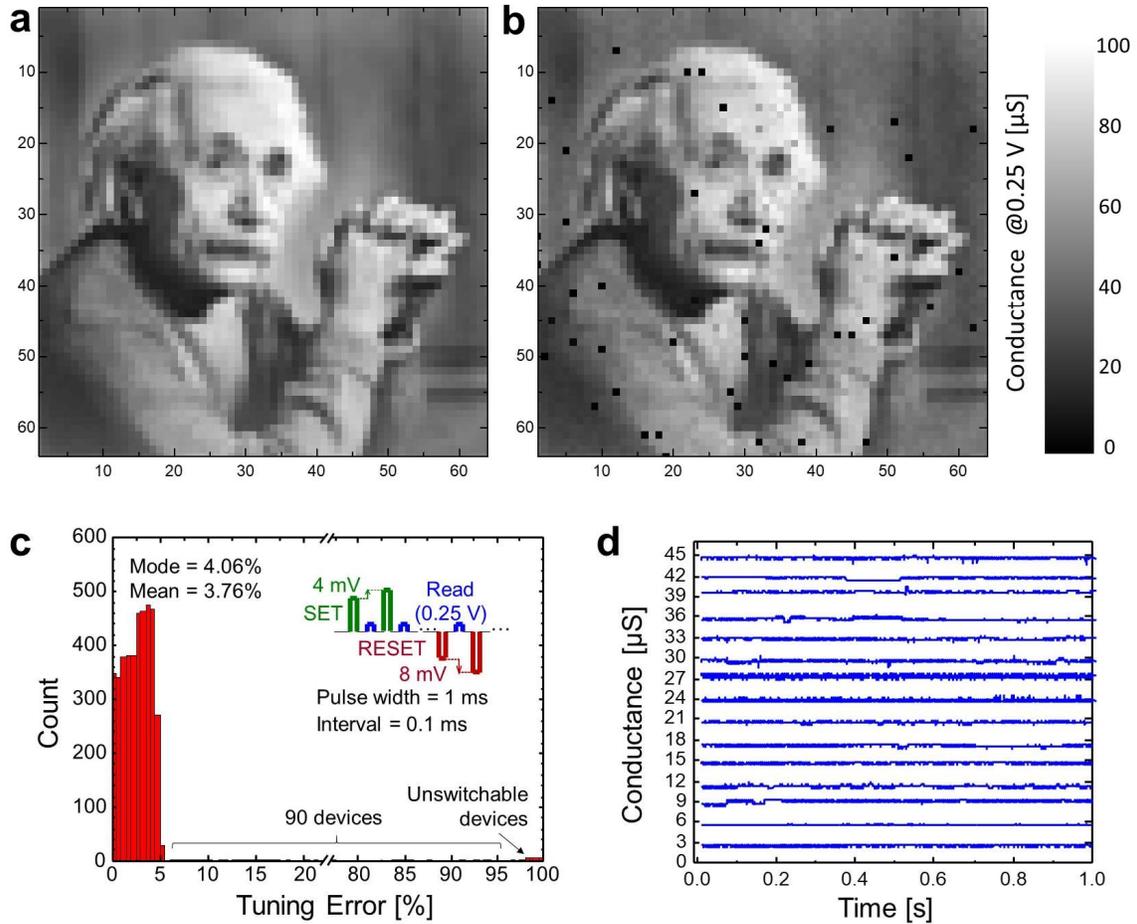

**Figure 3. Conductance tuning in the crossbar circuit.** (a) The desired device conductances in the range of 10 μS to 100 μS, which corresponds to the gray-scale quantized Einstein image and (b) their actual measured values after completing automated tuning with 5% error. (c) Corresponding tuning statistics. Inset shows details of the write-verify pulse sequence. Tuning error is defined as $100\times[I_{target}(0.25V)-I_{actual}(0.25V)]/I_{target}(0.25V)$. (d) Example of device tuning with 1% error to the different conductance levels equally spaced from 3 μS to 45 μS. All conductances are specified at 0.25 V.

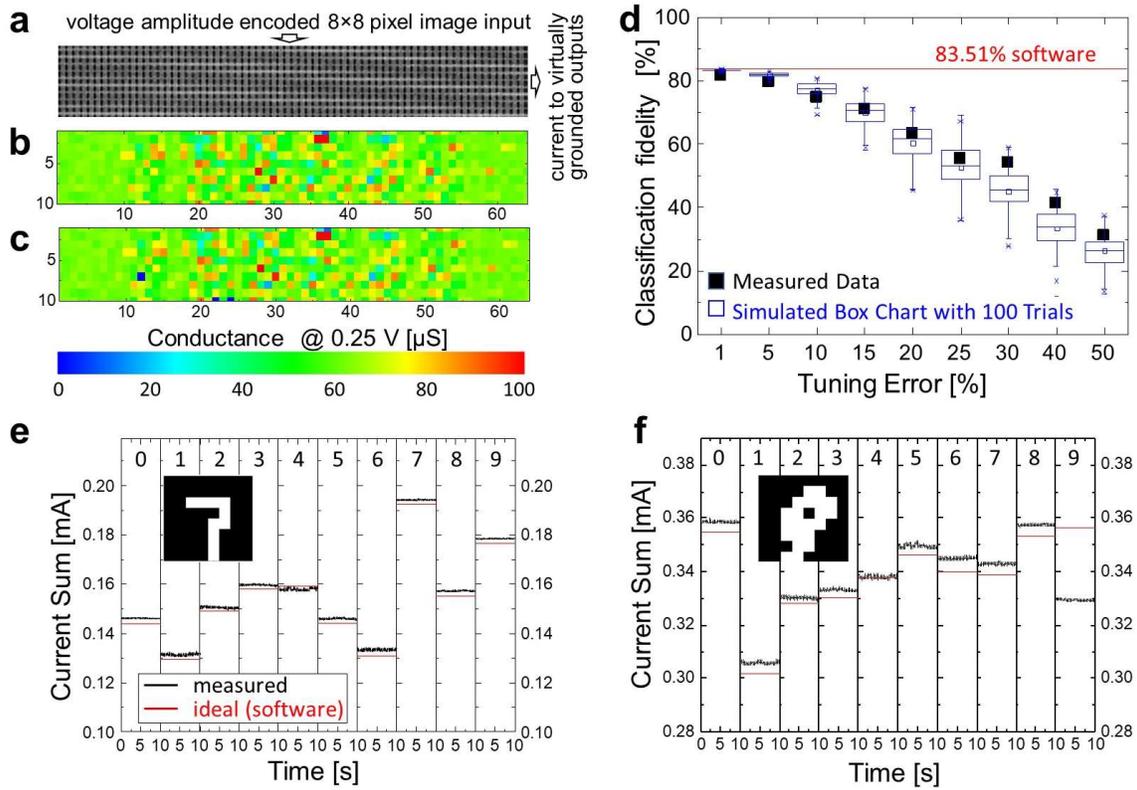

**Figure 4. Pattern classification of MNIST images.** (a) Portion of the crossbar circuit utilized in the implementation of 64×10 single-layer perceptron. (b) Examples of target and (c) actual conductances after tuning with 1% error. (d) Measured classification fidelity and its comparison with simulation results as a function of weight import accuracy. In each simulation trial, the weights were selected randomly from range of target value × [1 - tuning error, 1 + tuning error]. (e, f) Measured output currents for all ten outputs over 10 second interval for patterns '7' and '9' (shown in the corresponding insets) for the experiment with 1% tuning error.

## Supporting Information

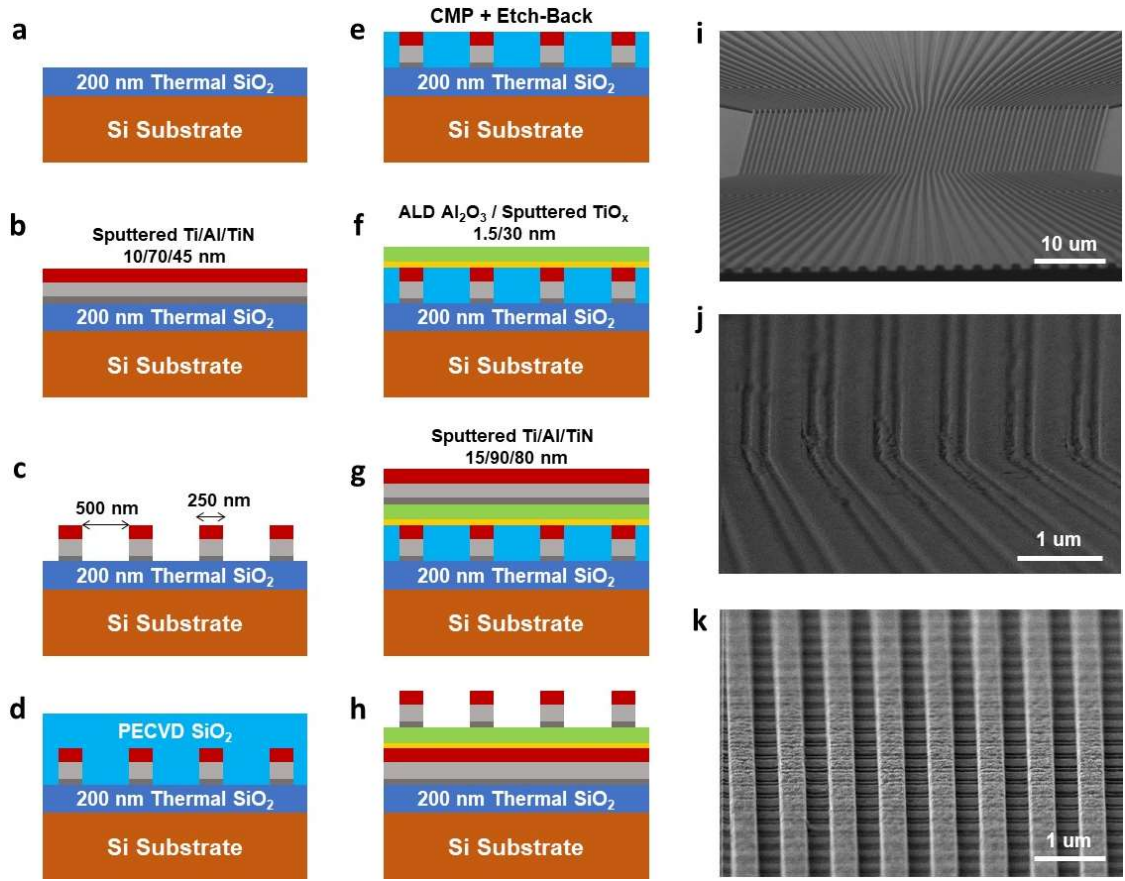

**Figure S1. Fabrication process flow.** (a-h) Growth and patterning process steps. For clarity, panel h shows cross-section turned out-of-plane by 90 degrees. Scanning electron microscopy images of (i) patterned bottom electrodes, (j) partially planarized bottom electrodes through chemical-mechanical polishing and etch-back, and (k) a fragment of completed crossbar array. All fabrication was performed at UC Santa Barbara's nanofabrication facility.

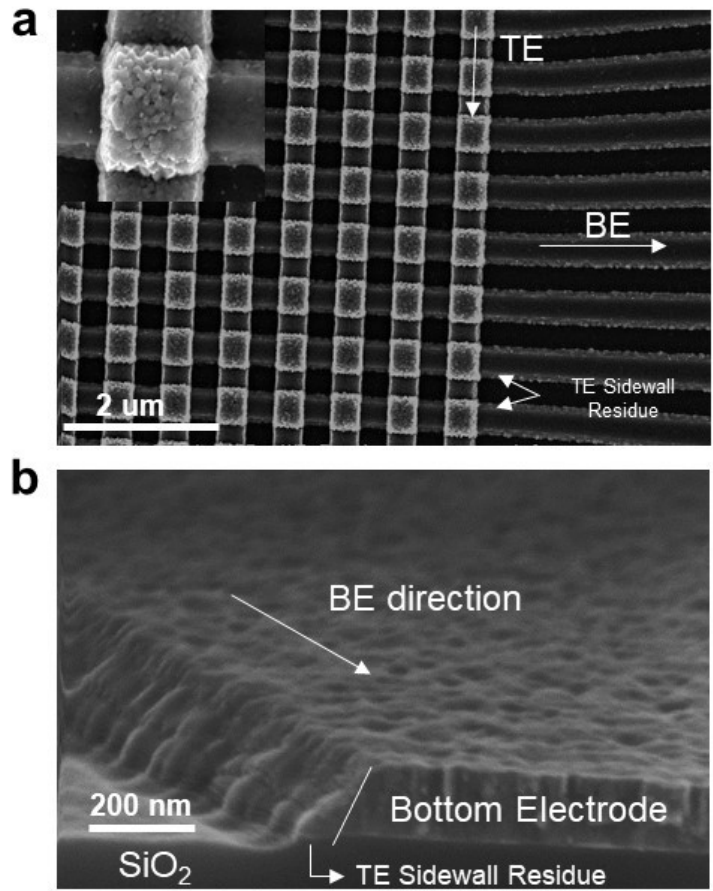

**Figure S2. Sidewall residue**. (a) A top and (b) cross-sectional scanning electron microscopy images of the crossbar circuit fabricated without planarization steps. Inset shows zoom-in on a crosspoint device. When the top electrodes (TEs) are patterned with etching process but without planarization step, a sidewall residue along the bottom electrodes (BEs) results in shortening of all TEs.

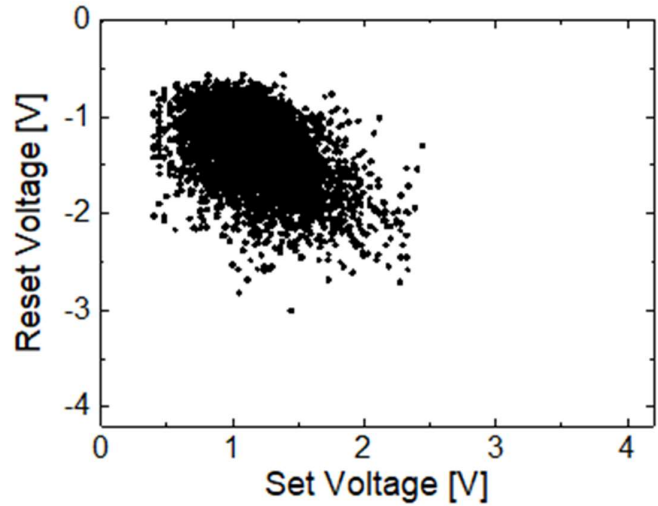

**Figure S3. Correlations in switching voltages.** Post-processed data from Fig. 2 showing significant correlations between set and reset switching threshold voltages.

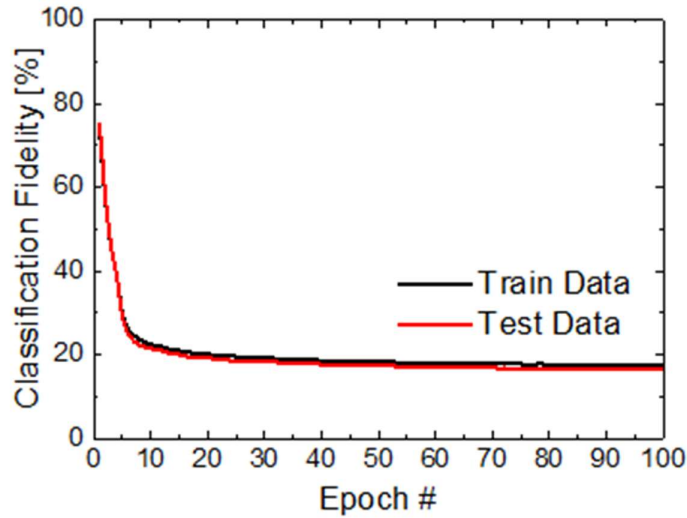

**Figure S4. Software-based training.** The training of a 64×10 single-layer perceptron classifier with rectify-linear neurons on 60,000 training and 10,000 test 8×8 down-sampled MNIST images, using conventional backpropagation algorithm with 0.01 learning rate, 100 of batch size, and 50% dropout rate. The network weights were clipped in [0, 1] range during training.